# Long Pulse EBW Start-up Experiments in MAST

V.F. Shevchenko[1, a], T. Bigelow[2], J.B. Caughman[2], S. Diem[2], J. Mailloux[1], M.R. O'Brien[1], M. Peng[2], A.N. Saveliev[3], Y. Takase[4], H. Tanaka[5], G. Taylor[6]

[1] CCFE, Culham Science Centre, Abingdon, OX14 3DB, UK

[2] Oak Ridge National Laboratory, Oak Ridge, Tennessee 37830, USA

[3] Ioffe Institute, Politekhnicheskaya 26, 194021 St. Petersburg, Russia

[4] University of Tokyo, Kashiwa 277-8561, Japan

[5] Graduate School of Energy Science, Kyoto University, Kyoto 606-8502, Japan

[6] Princeton Plasma Physics Laboratory, Princeton, New Jersey 08543, USA

[a] Corresponding author: vladimir.shevchenko@ccfe.ac.uk

**ABSTRACT**

The non-solenoid start-up technique reported here relies on a double mode conversion for electron Bernstein wave (EBW) excitation. It consists of the mode conversion of the ordinary mode, entering the plasma from the low field side of the tokamak, into the extraordinary (X) mode at a mirror-polarizer located at the high field side. The X mode propagates back to the plasma, passes through electron cyclotron resonance and experiences a subsequent X to EBW mode conversion near the upper hybrid resonance. Finally the excited EBW mode is totally absorbed at the Doppler shifted electron cyclotron resonance. The absorption of EBW remains high even in cold rarefied plasmas. Furthermore, EBW can generate significant plasma current giving the prospect of a fully solenoid-free plasma start-up. First experiments using this scheme were carried out on MAST [V. Shevchenko et al, Nucl. Fusion **50**, 022004 (2010)]. Plasma currents up to 33 kA have been achieved using 28 GHz 100kW 90ms RF pulses. Recently experimental results were extended to longer RF pulses showing further increase of plasma currents generated by RF power alone. A record current of 73kA has been achieved with 450ms RF pulse of similar power. The current drive enhancement was mainly achieved due to RF pulse extension and further optimisation of the start-up scenario.

[a] Corresponding author: vladimir.shevchenko@ccfe.ac.uk

## I. **INTRODUCTION**

Non-solenoid plasma current start-up is a critical issue in spherical tokamak (ST) research because of lack of space for a shielded central solenoid. Various techniques have been proposed and developed in order to avoid a central solenoid in future ST devices [1–5]. In this paper, we report on the electron Bernstein wave (EBW) start-up technique based on a 28 GHz gyrotron capable of delivering 300 kW power for up to 0.5 s. Experiments were carried out on the Mega Amp Spherical Tokamak (MAST) based at Culham Science Centre, UK [6]. MAST has 5 pairs of poloidal field coils inside the vacuum vessel as shown in fig. 1. Upper and lower parts of P2 – P5 coils carry current in the same direction and generate vertical magnetic field required for plasma shaping and equilibrium. Upper and lower parts of the P6 coils carry current in opposite directions and generate radial magnetic field providing vertical control of the plasma. This set of poloidal coils except for P3 was used in the start-up experiments described below. The central rod (CR) current generating toroidal magnetic field $B_T$ was set to be of 2 MA during these experiments. That gives a radial position of electron cyclotron (EC) resonance for 28 GHz ($B_T$ = 1 T) at 0.4 m.

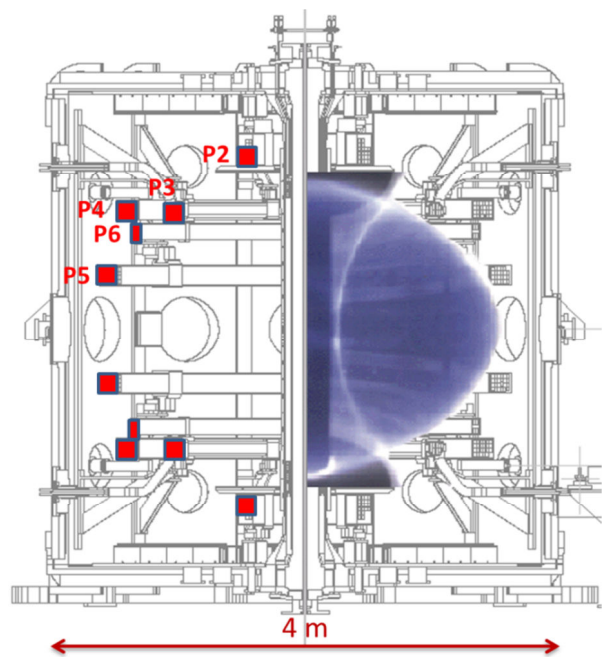

**Figure 1.** Location of poloidal field coils in MAST (left) and a typical plasma cross-section (right).

The plasma start-up method deployed here has been proposed in [7, 8] and first experiments were described in [1]. The method relies on the production of low-density plasma by RF pre-ionization around the fundamental EC resonance (ECR) with an ordinary (O) mode, incident from the low field side of the tokamak. Then the O mode is reflected back by a grooved mirror-polarizer incorporated in a graphite tile on CR. Polarization of the reflected beam is rotated by 90° to convert to the extraordinary (X) mode. The launched O-mode Gaussian beam is tilted to the midplane at 10° and hits the CR at the midplane as illustrated in fig. 2a

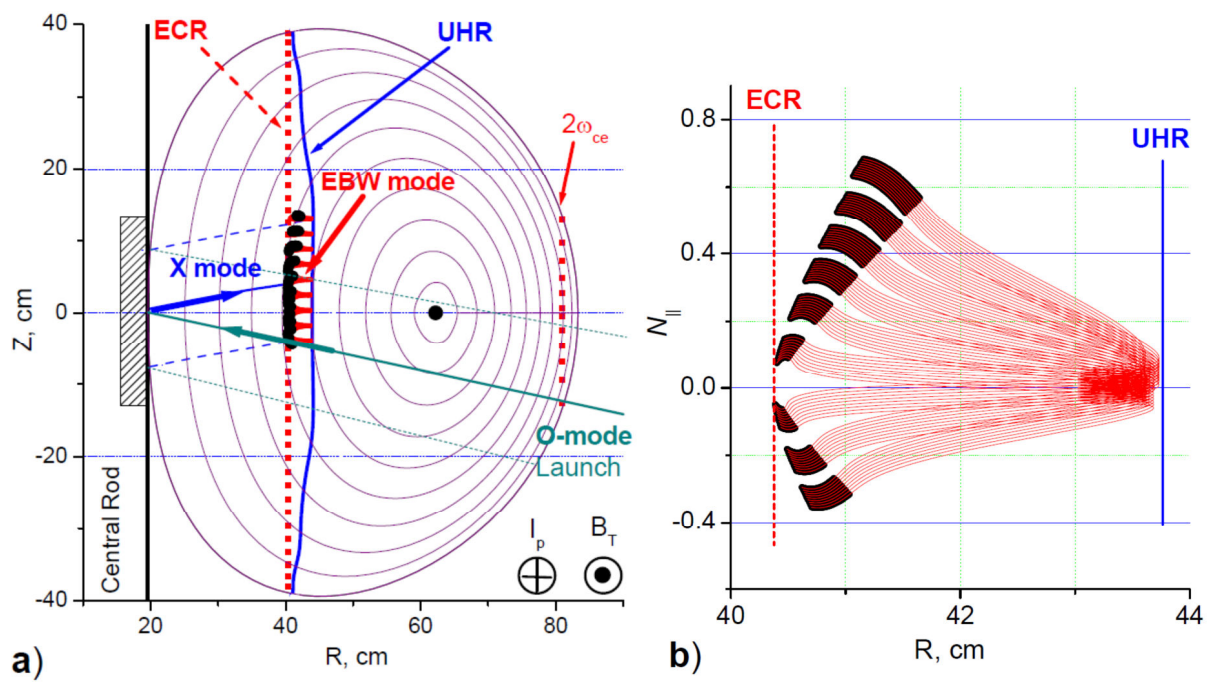

**Figure 2.** EBW assisted plasma start-up schematic with over plotted EBW ray-tracing results. Black colour at the end of EBW rays indicates the power deposition zone. **a)** poloidal cross-section: rays absorbed predominantly above the midplane; **b)** in the absorption zone EBW rays develop $N_{\parallel} > 0$ above the midplane and $N_{\parallel} < 0$ below. Note, about 5 times higher than usual density $n_e(0) = 1.5 \cdot 10^{18} \text{m}^{-3}$ is depicted. In this case the ECR and UHR layers are well separated while preserving the main features of EBW propagation.

The X mode reflected from the mirror-polarizer propagates back into the plasma and hits the ECR layer predominantly above the midplane. The X mode propagating nearly perpendicular to the magnetic field has very small absorption at the fundamental EC resonance within a wide range of plasma parameters typical for tokamak. Thus almost all the X-mode power passes through the ECR and experiences a subsequent slow X to EBW mode conversion (MC) near the upper hybrid resonance (UHR). Finally the excited EBW mode is totally absorbed before it reaches the ECR, due to the Doppler shifted resonance. Modelling shows that only a small fraction of the injected RF power (~2%) is typically absorbed as the O and X modes in start-up plasmas, while the main part is converted into the EBW mode. The absorption of EBW remains high even in cold rarefied plasmas. Indeed, one can estimate absorption using a simple analytical formula obtained within a global wave-dynamical treatment of MC processes [9]. It gives good agreement with numerical simulation within the range of parameters typical for start-up plasmas.

$$\tau = \frac{\pi}{2} \cdot R \cdot k \cdot \frac{\omega_{pe}^2}{\omega_{ce}^2}, \qquad (1)$$

where $\tau$ is an optical thickness, $R$ is a major radius of the plasma, $k = \omega_{RF}/c$ is a wave vector, $c$ is a speed of light $\omega_{pe}$ and $\omega_{ce}$ are a plasma frequency and an electron cyclotron frequency respectively. For the MAST start-up parameters it can be simplified even further:

$$\tau = 120 \cdot n_e^{18}, \qquad (2)$$

where $n_e^{18}$ is an electron density in unit of $10^{18} \text{m}^{-3}$. Finally the absorption coefficient $A$ can be estimated as usual:

$$A = 1 - e^{-\tau}. \qquad (3)$$

One can see that absorption is very close to 100% within a wide range of densities and it is independent of plasma temperature. Furthermore, EBW can generate significant plasma current (if EBW absorption is localised predominantly above or below the midplane to gain a directionality with respect to the magnetic field) during the start-up phase giving the prospect of a fully non-inductive plasma start-up [10].

The described schematic has natural limitations. The first limit comes from accessibility requirements. The plasma must be transparent for the O mode and the deflection of the O-mode beam due to refraction should not exceed the size of the mirror polariser at CR. That means that the plasma must be well under-dense for the RF frequency $\omega_{RF}$ injected into the plasma, i.e $\omega_{RF}^2 >> \omega_{pe}^2$, or equivalently $n_e <<$ $n_e^{cut-off}(\omega_{RF})$ with $n_e^{cut-off} = 9.7 \cdot 10^{18} \text{m}^{-3}$ for 28 GHz discussed here. The second limit comes from the lower density case. If the density is so low that the following inequality is valid:

$$\frac{\omega_{pe}^2}{\omega_{RF}^2} < \frac{T_e}{m_e c^2},$$ (4)

where $T_e$ is an electron temperature and $m_e$ is an electron mass, then the X-B MC does not take place and the X mode passes through the plasma [9]. During the RF breakdown phase this low density limit is usually quickly exceeded and the gas puff needs to be controlled to avoid the plasma reaching over dense condition.

The usual relativistic resonance condition must be satisfied by electrons near the fundamental EC resonance to provide efficient interaction with the waves,

$$\omega_{RF} = \frac{\omega_{ce}}{\gamma} + k_{||} v_{||},$$ (5)

where $\gamma = (1 - v_\perp^2/c^2 - v_{||}^2/c^2)^{-1/2}$ is the relativistic factor, $v_\perp$ and $v_{||}$ are the components of the electron velocity perpendicular and parallel to the magnetic field, and $k_{||} = N_{||} \cdot k$ is the component of the wave vector $k$ parallel to the magnetic field $B$ and $N_{||}$ is the corresponding refractive index. Even for start-up plasmas with relatively low electron temperature the Doppler downshift of ECR is important in the case of EBWs because while approaching the resonance they can develop very large $k_{||}$. In the presence of high power RF heating the electron distribution function may develop high energy tail then relativistic effects can also be important.

The EBW wave vector is essentially perpendicular to the UHR layer so $N_\perp$ dominates $N_{||}$ by about two orders of magnitude. The sign of $N_{||}$ determines whether EBW interacts with electrons moving along or opposite the magnetic field as follows from Eq. 5. The sign of $N_{||}$ is given by the projection of the wave vector $k$ on the local magnetic field near the UHR where EBW originated. In start-up plasmas usually $B_T >> B_P$ hence $k_{||} \approx k_T + k_P B_P/|B|$, where $k_T$ and $k_P$ are the toroidal and poloidal components respectively. At their origin EBWs have $k$ vector almost perpendicular to the UHR layer resulting in $|k_P B_P/|B|| >> |k_T|$, except for in the vicinity of the midplane. Therefore the sign of $k_{||}$ is mainly determined by the sign of $k_P \ B_P \ /|B|$. As seen in fig. 2b as EBW propagates towards the EC resonance, $k_\parallel$ is developed further due to poloidal plasma inhomogeneity. This effect determines a different sign of $k_\parallel$ above and below the midplane. Moreover, $k_P \ B_P \ /|B|$ changes sign if the plasma current is reversed and remains unchanged if the toroidal field is reversed [10]. EBW rays propagating close to the midlane usually do not contribute significantly to the current drive because their $N_\parallel$ experiences oscillations between small positive and negative values which result in a close to zero net current after averaging over the rays around the midplane [11, 12]. A detailed theoretical study of EBW propagation in tokamak plasma can be found in [13] and in particular near the midplane [14 ].

## II. LONG PULSE EXPERIMENTS

The original hardware was significantly upgraded since first experiments [1] in collaboration with the Oak Ridge National Laboratory, USA. A short pulse gyrotron has been replaced with a 28 GHz 200 kW CW tube which can deliver up to 300 kW in a pulsed mode with the pulse duration up to 0.5 s. The original 30 m transmission line based on smooth 40 mm circular $TE_{01}$ waveguides has been upgraded with 89 mm $HE_{11}$ corrugated waveguides resulting in improvement of total transmission efficiency up to 75%. The entrance into the vacuum vessel and final focusing of the RF beam remained unchanged. It was done quasi-optically through a fused silica vacuum window with a clear aperture of 55 mm.

The quasi-optical launcher consists of 2 ex-vessel and 3 in-vessel ellipsoidal mirrors. The first 2 mirrors provide beam waist of 15 mm at the vacuum window to minimize reflection. The last launching mirror with an aperture of 230 mm focuses the beam to the grooved mirror-polarizer (see Fig. 2a) which is machined on the surface of the graphite tile. In-vessel low-power RF tests showed that the beam measured at the CR has an effective (electric field e-fold) radius of about 80 mm which is well within the grooved area of $250{\times}250 \ mm^2$. The high power experiments have also confirmed good focusing of the launched beam as the "hot" spot observed with an infrared camera was well centred at the mirror-polarizer and its angular dimensions agreed with the beam radius of 80 mm. Total resistive losses on the in-vessel mirrors were estimated to be less than 5%.

The reflected X-mode beam experiences very small vertical divergence. As seen in Fig. 2a its vertical effective radius remains almost unchanged on its way towards ECR. Typically plasma is well under-dense ($\omega_{RF}^2 >> \omega_{pe}^2$) during EBW start-up so the UHR layer is separated from ECR by only a few centimetres. For illustration purpose Fig. 2 shows the case where separation of the resonance layers was enhanced by additional gas puffing which caused an increase of the central density up to $1.5 \cdot 10^{18}\,\mathrm{m}^{-3}$. Such density is about five times higher than typical values so that the separation of layers is visible. The vertical size of the RF beam cross-section with UHR depends on plasma density but usually it is very similar to that at the CR. One can estimate a maximum electric field achievable at the ECR. For 100 kW of injected RF power we have $E_{RF} \approx 600$ V/cm for the O mode and about $\sqrt{2}$ times smaller for the X mode due to divergence in the toroidal plane after reflection from the cylindrical CR. A critical field for 28 GHz breakdown in deuterium in the present MAST geometry is about 700 V/cm [15]. None of the above modes can produce breakdown at power levels of 100 kW or below. However their presence increases the density of free electrons around the ECR layer making possible EBW excitation (see Eq. 4) at the UHR. The electric field of the EBW mode experiences strong (typically by factor of ~10) enhancement near the UHR which is sufficient for production of well sustained breakdown at the RF power levels much lower than 100 kW.

Unfortunately the RF power injected into the MAST vessel was limited by arcing at the focusing optics near the MAST vacuum window. So the experiments reported here extend previous results only in terms of pulse duration with about the same level of RF power delivered into the vessel. The power within the vessel was monitored with a stray radiation detector (SRD) which was part of the EC radiometer discussed in a section IV. The SRD generated a signal proportional to the power density bouncing inside the vessel. Typically after the RF power onset it showed a signal which was gradually decreasing to about half of its initial value during a time interval from 5 to 10 ms depending on RF power, gas puff and vertical magnetic field. This interval was identified as a free electron density build-up phase. Then the SRD signal dropped suddenly and exponentially down to less than one tenth of its previous value within 0.1 - 0.2 ms indicating the onset of the EBW breakdown phase. From this moment a strong growth of plasma density, EC emission and plasma current was observed. This behaviour of signals was reproduced in the whole range of experiments discussed below providing evidence that typically more than 95% of the injected power was absorbed by the plasma. Single pass absorption close to 100% was reported for other start-up experiments with high field side X-mode launch [16].

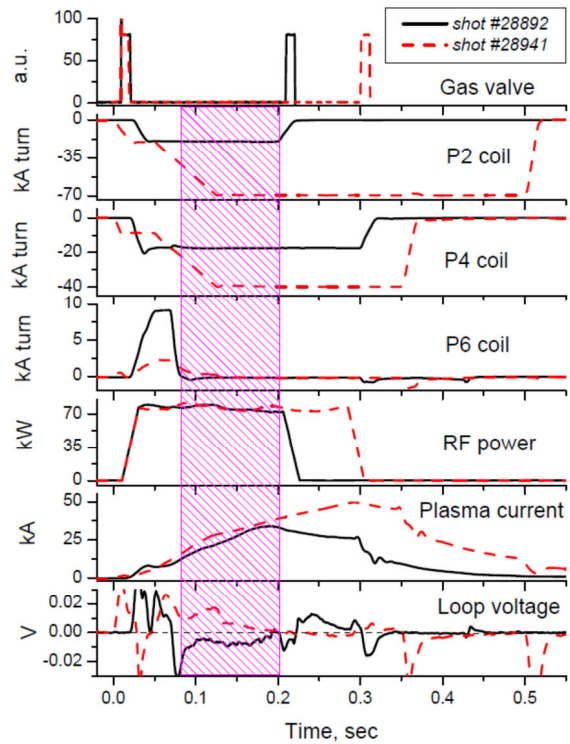

**Figure 3.** EBW start-up experiments in constant Bv. Experimental waveforms of 2 shots with different Bv values. The shaded area illustrates negative loop voltage generation in the shot #28892.

First let us consider EBW start-up experiments in a constant vertical field Bv. Waveforms are shown in fig. 3. A gas valve puffs the gas just before the RF pulse. In the shot #28892 Bv is produced by the P2 and P4 coils and it reaches its stationary value in 15 ms after RF onset. These coils generate Bv of about 5 mT at the ECR location near the midplane. The resulting Bv field has a barrel shape with a minimum at the midplane. Then P6 coils produce about 15 cm up-down vertical shifts of the Bv minimum, which do not generate any loop voltage or pre-ionisation but help to form closed flux surfaces [1]. Plasma current $I_p$ monotonically increases from 0.1 s until 0.2 s when the P2 current is switched off. After that $I_p$ shows a gradual decrease and then reduces rather quickly when the P4 current is switched off. The shaded area indicates the time interval when all external magnetic fields were kept constant.

However a negative loop voltage is clearly seen in fig. 3 during this stationary phase 0.08 – 0.2 s. It achieves its most negative value of -10 mV during 0.14 – 0.17 s and then gradually goes to zero at 0.19 s. There are no other sources of toroidal electric field apart from increasing Ip driven by the RF power and the highest Ip ramp-up rate is seen during the 0.14 - 0.17 s interval. The plasma current reaches its saturation at 0.19 s because constant Bv is no longer sufficient to confine the plasma from expansion.

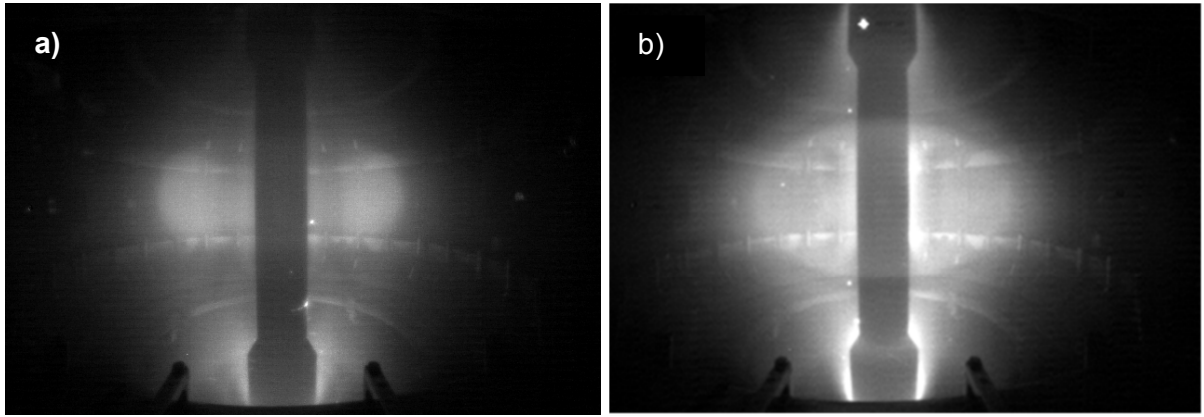

**Figure 4.** Plasma images taken during EBW start-up with constant Bv, shot #28941: a) at 0.15 s shortly after closed flux surfaces formation; b) at 0.28 s close to the end of RF injection.

In the second shot #28941 shown in fig. 3 Bv is partly set before the RF pulse. Then Bv is ramped up to a higher value during P6 vertical kicks. Then all fields remain constant from 0.12 s through the whole duration of the RF pulse until 0.29 s. As in the previous shot Ip monotonically increases until the end of RF injection. In both shots plasma current reaches the same value at 0.19 s but in the second one it was achieved with smaller radial field but higher Bv ramp-up. The smaller radial field is required to produce the same vertical shift because Bv is smaller. Monotonic Ip growth until the end of RF pulse indicates that saturation has not been achieved with this value of Bv. Immediately after the end of the RF injection Ip starts to decrease and at a rate which then increases dramatically when the P4 coil current is switched off and the plasma loses equilibrium.

Fig. 4 illustrates different phases of plasma evolution during start-up in constant Bv. The first picture (fig. 4a) was taken at 0.15 s straight after closed flux surfaces formation and the second one (fig. 4b) at 0.28 s close to the end of the RF pulse when plasma has expanded both vertically and radially. It was shown experimentally that the most efficient way to achieve higher plasma current and keep plasma in equilibrium is to apply Bv ramp up with such a waveform that keeps the loop voltage close to zero. Waveforms of plasma current generated by RF pulses of different duration with similar waveforms of Bv ramp up are shown in fig. 5. Here RF power of the same level is injected into the plasma for 320 ms, 400 ms and 440 ms. It is clear from this figure that the plasma current drops by ~20% within ~30 ms of the end of the RF pulse and then remains almost constant until the end of the Bv ramp up.

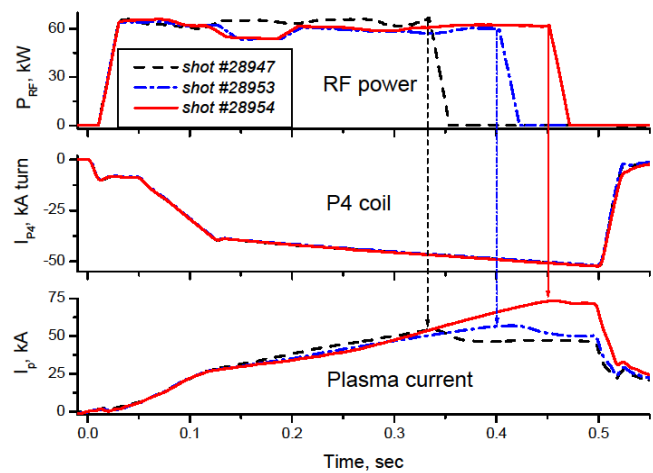

**Figure 5.** Effect of the RF pulse length with the ramping-up Bv.

It was concluded in [1] that the plasma current is predominantly carried by supra-thermal electrons with energy of several tens of keV. This estimate was based on plasma current decay measurements after the end of RF injection. More accurate measurements of fast electrons are discussed in the section IV.

Such electrons must be almost collisionless in the plasma with the density of $3 \cdot 10^{17}$ m$^{-3}$ so they should not slow down after the RF drive is switched off. There is also a fraction of the plasma current carried by slower electrons which will decay faster than the supra-thermal fraction providing the observed ~20% decrease of the total plasma current. The supra-thermal part of the current remains constant until the end of Bv ramp due to a small loop voltage generated by Bv ramp-up. If Bv is kept constant the remaining supra thermal Ip decays typically with a time constant of about 0.5 seconds.

The line averaged plasma density in these experiments was about $3 \cdot 10^{17}$ m$^{-3}$ as estimated from interferometric measurements. Unfortunately detailed measurements of electron temperature and density profiles were not possible because the plasma density was below the Thomson scattering sensitivity limit. An approximate model of the start-up plasma was developed for modelling of the start-up experiments. TS measurements were often possible straight after a second gas puff at the end or after RF injection as depicted in Fig. 3. Typically during gas puffing the line integrated plasma density showed rapid increase for about 10 ms followed by a gradual decrease with a time constant of about 50 ms. It was assumed that the energy loss was small compared with the stored energy during the rapid density rise. The high energy fraction of electrons remained collisionless so only the energy stored in the bulk plasma was redistributed between existing electrons and electrons produced by ionisation without significant distortion of the profiles. Energy loss through ionisation was ignored. Typically after the gas puff the TS measured relatively flat profiles with central values of $n_e(0)$ within 0.8 - $1.2 \cdot 10^{18}$m$^{-3}$ and $T_e(0)$ within 60 - 80 eV with some peaking around ECR radius R = 0.4 m up to 140 eV. The increase of line integrated density was about factor of 3 for 10 ms gas puff pulses. Taking averaged values of these figures it was concluded that plasma profiles before gas puffing were flat with $n_e(0) \approx 3 \cdot 10^{17}$m$^{-3}$, $T_e(0) \approx 210$ eV with a localised peak in temperature of 420 eV around R = 0.4 m. These scaled profiles were used to conduct beam-tracing [17] for O and X-mode propagation followed by EBW ray-tracing [18] and Fokker-Planck simulations [19] discussed in the section III.

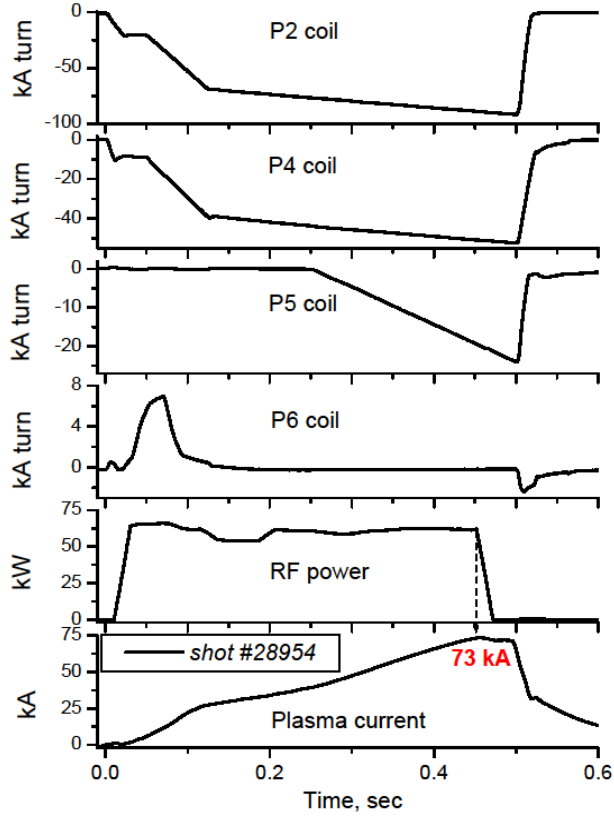

**Figure 6.** Waveforms of the shot #28954 with a record plasma current of 73 kA achieved in EBW start-up.

A record current of 73 kA has been achieved by optimisation of Bv waveforms as illustrated in fig. 6. RF power of about 60 kW with 0.44 s pulse duration was injected into the MAST vessel. Three pairs of poloidal field coils were used to provide optimised shaping and the Bv ramp-up rate. P5 coils were energised at 0.25 s to introduce additional shaping to the plasma which is favourable at higher Ip. The plasma current reached the value of 73 kA at the end of RF pulse and it would probably have continued to rise further if RF power had been available. A total efficiency of 1.2 A/W has been achieved with this start-up scenario. To our knowledge efficiencies close to 1 A/W were previously reported only for start-up scenarios based on lower hybrid waves [20]. For comparison of different experiments a dimensionless current drive (CD) efficiency is more commonly used [21]:

$$\zeta = \frac{e^3}{\varepsilon_0} \frac{n_e I_{CD} R}{P_{RF} k T_e} \cong 33 \frac{n_{20} I_A R_m}{P_W T_{keV}} \approx 0.16, \qquad (6)$$

where $n_{20}$ is the local density in units of $10^{20}$ m$^{-3}$, $I_A$ is the driven current in A, $R_m$ is the major radius in meters, $P_W$ is the RF power in W, and $T_{keV}$ is the local temperature in keV. This result is close to the dimensionless efficiency obtained in short pulse EBW start-up experiments [1].

The plasma image taken at the end of the RF pulse is shown in fig. 7. The plasma has a clear toroidal structure with elongated poloidal cross-section. Unfortunately due to low density in this shot Thomson scattering measurements of plasma density and temperature profiles were not possible.

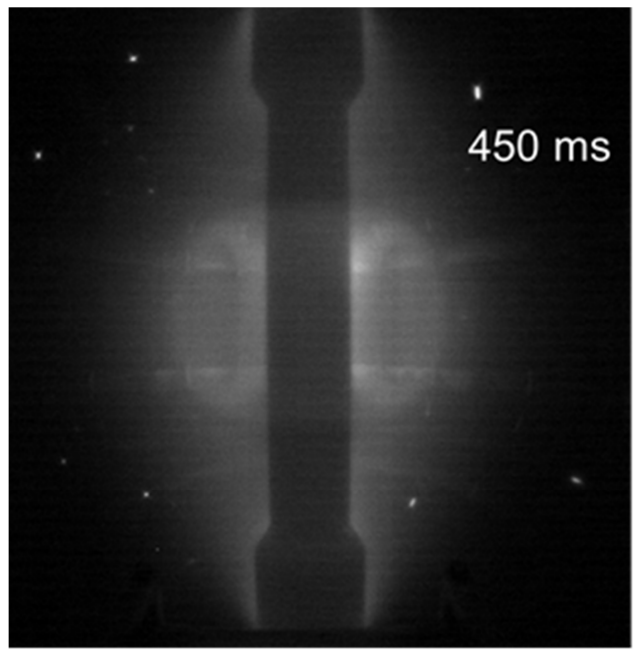

**Figure 7.** Plasma image at the end of RF pulse with the maximum plasma current of 73 kA.

As mentioned above the experiments presented here were constrained by the arcing limiting RF power injected into MAST. However in relatively short pulses (< 0.2 s) it was possible to inject RF power exceeding 100 kW. On the other hand it was found experimentally that the lowest power level which still reproduced the above start-up scenario is about 40 kW. Due to the fact that the higher power shots were limited in duration, a time slice available in the majority of experiments has been chosen for comparison. That is the moment when closed flux surfaces were formed and the initial Bv ramp-up and P6 vertical kick were completed. For the majority of shots this happened after 150 ms so the time at 160 ms was chosen for scaling. The results are depicted in fig. 8.

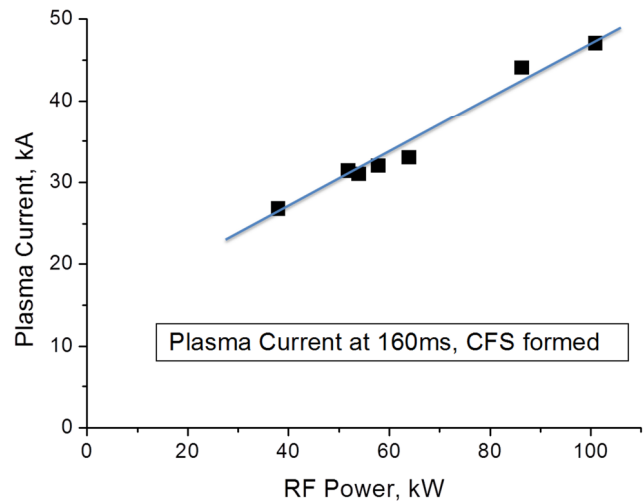

**Figure 8.** Scaling of start-up current measured after closed flux surfaces formation against injected RF power.

Interestingly, all experimental points fit a linear dependence of generated plasma current versus RF power injected into the plasma. At least this dependence seems to be valid within the range of RF power available in the experiments reported in this paper. In the present launcher set-up the RF breakdown occurs at power levels exceeding ~10kW. So the curve should start at the 10kW point followed by a stronger than linear dependence until it gradually joins the line presented in fig. 8.

## III. EXPERIMENTS WITH VERTICAL MODULATION

It was predicted in [11] and experimentally confirmed in [10] that the EBW power deposition location with respect to the plasma midplane is crucial for the EBW current drive (CD) direction. A special experiment was conducted to verify the CD mechanism responsible for the current generation in the start-up scenario under discussion. To change the vertical localisation of EBW power deposition within the plasma is impossible with the existing launching system because it was designed with reference to the mirror–polariser position. The entire RF launching system was carefully pre-aligned to provide efficient mode conversion on the mirror-polariser which is fixed at the midplane on the central rod. However it is possible to move the plasma midplane with respect to the machine midplane by applying radial magnetic fields produced by P6 coils. That would allow an experimental test of the effect of the EBW power deposition location on CD efficiency and direction.

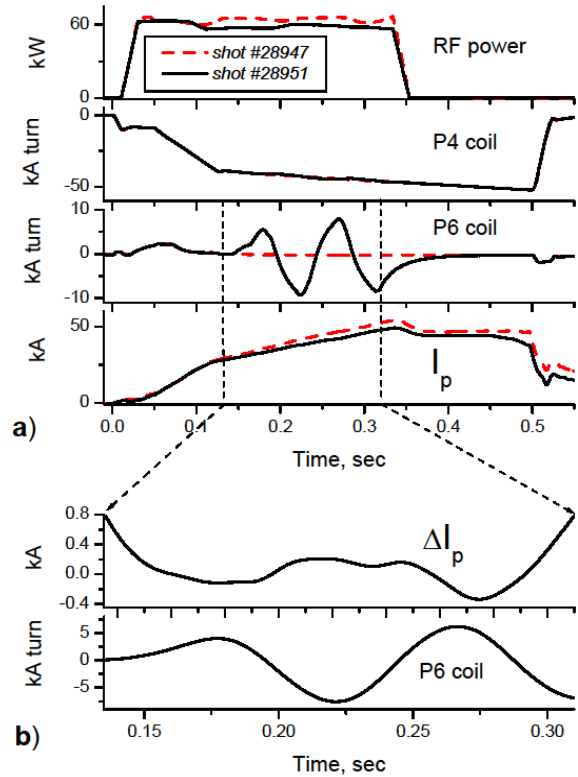

**Figure 9.** Waveforms of the experiment with plasma vertical modulation: **a)** reference shot #28947 and test shot #28951 waveforms; **b)** differential plasma current $\Delta$Ip = Ip$^{test}$-Ip$^{ref}$ and P6 coil current.

The experiment was conducted as follows. A reference shot was produced by 60 kW 350 ms RF injection with the same Bv waveforms as in fig. 6 apart from P5 which was disabled in these experiments. Then several test shots were fired with two period near sinusoidal modulations applied to the radial field coils P6 starting at 0.15 s as shown in fig. 9a. This sort of modulation produced vertical up and down plasma shifts of about +/-15 cm which were clearly visible with a fast video camera. Plasma current modulations induced by vertical shifts are barely visible on the waveforms. However, after subtraction of the plasma current from the reference shot and removing an offset the effect of modulation becomes clear as shown in fig. 9b with the zoomed-in differential signal. The plasma current decreases when the plasma moves up (positive P6 current swing) and increases when the plasma moves down (negative P6 current swing). That agrees well with the changes of sign of $k_\parallel$ as explained below.

Such behaviour can be understood from the following consideration. In a non-disturbed position shown in fig. 2 all EBW power is absorbed above the midplane. This power produces CD in the co-direction. When the plasma was moved up while the RF beam remained fixed some fraction of the EBW power will be deposited below the plasma midplane causing CD in a counter-direction. That will result in a reduction of the total CD and hence a reduction in plasma current. Conversely when plasma moves down the fraction of the power deposited above the plasma midplane becomes larger resulting in the increase of plasma current. These experiments qualitatively prove the EBW CD mechanism in this start-up scenario.

The MAST vessel (see Fig.1) is very large in comparison with the plasma dimensions during start-up. Thus with ±15 cm vertical shifts of the plasma one should expect very little or no effect at all on plasma shape or confinement. Magnetic reconstruction and plasma images do not show any visible changes. Therefore all effects can be attributed mainly to the variation of the power deposition with respect to the midplane.

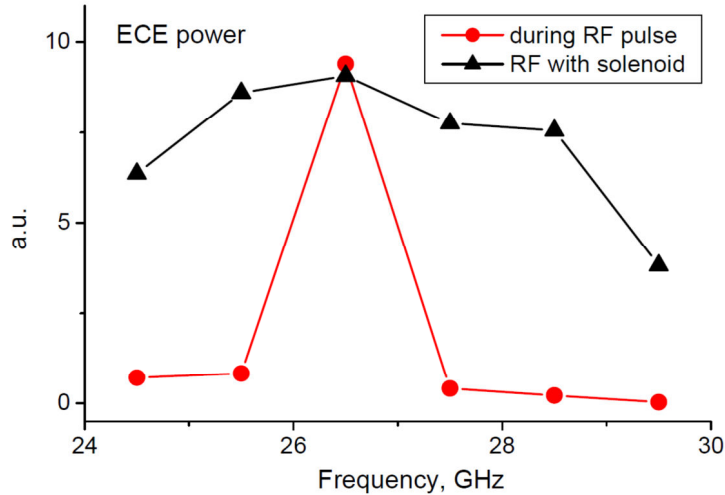

**Figure 10.** ECE spectra measured during EBW start-up in constant Bv with RF injection only (red circles) and with RF injection and with a limited solenoid assist (black triangles).

Ray-tracing coupled with relativistic Fokker-Planck modelling was conducted using assumptions about $n_e$ and $T_e$ profiles as discussed in the previous section. A simple, but sufficiently accurate, kinetic Fokker–Planck estimate of bootstrap effects is used. For the non-disturbed plasma position with the $T_e$ profile without the peak modelling predicted: 1.2% of total power was absorbed as the O mode, 0.9% as the X mode and the whole remaining power was absorbed as the EBW mode. Total steady-state current generated by 60 kW of RF power was 54 kA from which only 1.7 kA was estimated as a bootstrap current and no current was driven by O and X modes. For the profile with the peak of $T_e$ = 420 eV at $R$ = 0.4 m modelling predicted about 100 kA of driven current with the bootstrap fraction remaining below 3 kA. For the 15 cm up-shifted plasma the modelling predicted similar values of current but in a counter direction. It must be noted that the predicted currents are the steady-state currents and they require a so-called 'turn-on' time to achieve predicted values. That point will be discussed in the next section.

## IV. ECE FROM START-UP PLASMA

Another experiment conducted to aid understanding of the physics of EBW start-up was measurement of fundamental EC emission (ECE) from the plasma. A radiometer was receiving O-mode polarised ECE with the antenna viewing direction close to perpendicular ($N_\parallel \sim 0$) to the magnetic field. The radiometer frequency coverage was from 24.5 GHz to 29.5 GHz with the gyrotron frequency filtered out within 28±0.1 GHz by a notch filter with -60 dB suppression. Radiometer channels had a bandwidth of 0.5 GHz measured at -3 dB. In order to avoid radiometer saturation all measurements discussed below were obtained with a 10 dB attenuator installed at the antenna feed in comparison with routine plasma operation where no attenuation is required.

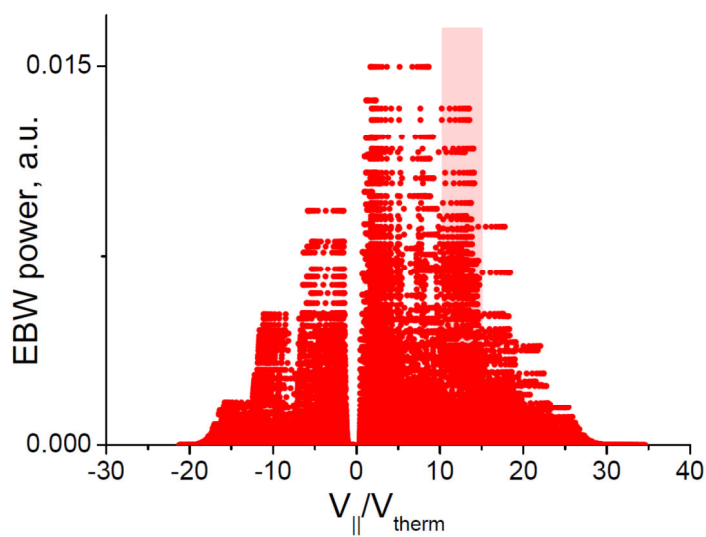

**Figure 11.** Relative power absorbed by electrons with particular $v_\parallel$ plotted for all EBW rays. The difference between power absorbed by electron with positive and negative $v_\parallel$ is responsible for the net generated current.

Results are summarised in fig. 10. It shows two ECE spectra. The data shown in red circles was taken during a constant Bv period when no positive loop voltage is present in the plasma. There is a clear maximum in the spectrum at 26.5 GHz which is down-shifted from the gyrotron frequency of 28 GHz by 1.5 GHz. For comparison the data plotted in black triangles shows the spectrum measured during RF injection with a limited solenoid assist. The solenoid produced a loop voltage of 0.2 V for only 20 ms which was sufficient for generation of high energy electrons. The measured spectrum does not show any clear peaks or valleys around 28 GHz and the ECE intensity is very close to the maximum of the red circle spectrum. This proves that the notch filter did not affect the neighbouring channels and the radiometer had a flat response over the range of frequencies. Obviously in both cases ECE is emitted by supra-thermal electrons. However the peaked spectrum is produced by electrons localised near the Doppler downshifted ECR layer and accelerated by RF waves alone while the flat spectrum is produced by electrons from the entire plasma cross-section and accelerated predominantly by the applied loop voltage.

For the case of $N_\parallel \sim 0$ equation 5 is simplified to $\omega_{ECE} = \omega_{ce} \,/\, \gamma$ so the relativistic factor $\gamma$ can be straightforwardly estimated if we assume that EC emission originated from the layer coincident with the ECR at 28 GHz. For the peak frequency of 26.5 GHz it gives $\gamma = 1.06$, hence $v^2/c^2 \approx 0.1$ and $m_e \cdot v^2/2 \approx 27$ keV. That is the energy which electrons acquire from the RF field in the ECR layer. The ray-tracing modelling shows that EBWs develop $N_\parallel$ within the range of 0.3-0.5 as they approach ECR. Simulations show that EBWs can generate a fast electron tail with energy of 25-50 keV which is predominantly responsible for the plasma currents observed in the above experiments. Only $1.2 \cdot 10^{16}$ of 27 keV electrons are required to carry the whole 73 kA current which is about 4% of the total number of electron in the start-up plasma, Fig. 11 represents the relative power absorbed by electrons with particular $v_\parallel$ along EBW rays. Contributions from all EBW rays were integrated over the volume where absorption was strong. The effective normalised radius for the maximum absorption in this particular case was about 0.6. The difference between power absorbed by electrons with positive and negative $v_\parallel$ is responsible for the net generated current. Apparently the shaded range of velocities $11 < v_{\parallel}/v_{therm} < 15$ makes a predominant contribution into the net current ($v_{therm}$ is taken for $T_e$ = 210 eV). The plasma current turn-on time $\tau_{t-o} \approx [6w_1^2\sqrt{w_2 - w_1}]\tau_{ee}$ is about 200 ms for $T_e$ = 210 eV and about 500 ms for $T_e$ = 420 eV which is close to the experimentally observed values. Here we used the same notation as in [22] where $w_1 = 11$ and $w_2 = 15$ are the spectrum edges in the velocity space normalised by thermal velocity and $\tau_{ee}$ is the electron-electron collision time of the bulk plasma. In the experiments with vertical modulation the plasma was shifted up for only about 15 ms so the difference in current gained during this time must be scaled for comparison with the steady-state values. Appling the scale factor gives the difference in current -0.4 kA·500/15 = -13 kA which is still much lower than the predicted value of -54 kA. This discrepancy can be attributed to the fact that simulations started with Maxwellian distribution while in the experiment the counter-CD was applied to the electron distribution which had already developed a high energy tail due to the previous co-CD. Obviously, more detailed measurements in a wider parameter range are necessary in order to clarify all the physics involved in EBW start-up. There is no evidence of run-away electrons in the plasma generated by the described method.

## V. CONCLUSIONS

Improvements, upgrades and associated testing of the high power 28 GHz system used for the MAST EBW start-up have been performed as collaborative research between ORNL and CCFE. RF power injected into the vessel was of the same level as in previous experiments [1] but the RF pulse length was increased by a factor of 5 (up to ~0.5 s). These long pulse EBW start-up experiments on MAST have more than doubled the previously achieved plasma current levels. Closed flux surfaces were clearly observed indicating that tokamak like equilibrium was established during the EBW start-up phase. The response of the plasma current to vertical modulation of the plasma midplane demonstrated that the current was driven by EBWs. However modelling predicts a smaller total current generated by EBWs but a stronger response of the plasma current to the vertical modulation than observed in the experiments. A power scan indicates that the plasma current generated after closed flux surfaces formation depends linearly on the RF power injected. ECE measurements during EBW start-up support the hypothesis that plasma current is predominantly carried by supra-thermal electrons with energies about 27keV.

**Acknowledgements**

This work was funded partly by the RCUK Energy Programme under grant EP/I501045 and the European Communities under the contract of Association between EURATOM and CCFE. To obtain further information on the data and models underlying this paper please contact PublicationsManager@ccfe.ac.uk. The views and opinions expressed herein do not necessarily reflect those of the European Commission.

**References**

1. V. Shevchenko et al, Nucl. Fusion **50,** 022004 (2010)

2. C.B. Forest et al, Phys. Rev. Lett. **68** 3559 (1992)

3. R. Raman et al, Phys. Rev. Lett. **90,** 075005 (2003)

4. A. Ejiri et al, Nucl. Fusion **46,** 709 (2006)

5. T. Maekawa et al, Nucl. Fusion **52,** 083008 (2012)

6. B. Lloyd et al, Nucl. Fusion **43**, 1665 (2003)

7. V. Shevchenko et al., Proc. 13th Joint Workshop on ECE and ECRH, 255 (2004)

8. V. Shevchenko et al, Fusion Sci. Technol. **52,** 202 (2007)

9. V. Petrillo, G. Lampis and C. Maroli, Plasma Phys. Cont. Fusion **29**, 877 (1987)

10. V. Shevchenko et al, Phys. Rev. Lett. **89,** 265005 (2002)

11. C.B. Forest et al, Phys. Plasmas **7,** 5**,** 1352 (2000)

12. J. Urban et al, Nucl. Fusion 51 (2011) 083050

13. A.D. Piliya and E.N. Tregubova, Linear conversion of electromagnetic waves into electron Bernstein waves in an arbitrary inhomogeneous plasma slab, Plasma Physics and Contred Fusion, 47 (2005) 143-154

14. A.D. Piliya, A.Yu. Popov and E.N. Tregubova, Electron Bernstein waves in the mid-plane region of a spherical tokamak, Plasma Phys Control Fusion, 47 (2005) 379-394

15. M. Gryaznevich, V. Shevchenko and A. Sykes, Nucl. Fusion, 46 (2006), S573

16. D.R. Whaley et al, Nucl. Fusion, 32, No.5 (1992) 757

17. A.N. Saveliev, Plasma Phys Control Fusion, 51 (2009) 075004

18. A. D. Piliya, A. N. Saveliev and E. N. Tregubova, Proc. 30th EPS Conference on Contr. Fusion and Plasma Phys., 2003, 27A, P-3.203

19. M.R. O'Brien et al. in Proc. IAEA Technical Committee Meeting on Advances in Simulation and Modelling of Thermonuclear Plasmas, Montreal 1992, p.527

20. F. Jobes et al, Phys. Rev Lett. **52**, 12, 1005 (1984)

21. T. C. Luce et al, Phys. Rev. Lett. 83 (1999) 4550

22. N.J. Fisch, Reviews of Modern Physics 59, 1, 175 (1987)